\documentclass[pre,twocolumn,a4paper]{revtex4}

\usepackage{amsfonts}
\usepackage{amsmath}
\usepackage{amssymb}
\usepackage[dvips]{graphicx}
\usepackage[latin1]{inputenc}
\usepackage{array}

\begin{document}

\date{29 Marts 2007}
\author{Misha Marie Gregersen}
\author{Laurits Højgaard Olesen}
\author{Anders Brask}
\author{Mikkel Fougt Hansen}
\author{Henrik Bruus}
\title{Reversed flow at low frequencies in a
microfabricated AC electrokinetic pump}
\affiliation{
 MIC -- Department of Micro and Nanotechnology,
 Technical University of Denmark\\
 DTU bldg.~345 east, DK-2800 Kongens Lyngby, Denmark}

\begin{abstract}
Microfluidic chips have been fabricated to study electrokinetic
pumping generated by a low voltage AC signal applied to an
asymmetric electrode array. A measurement procedure has been
established and followed carefully resulting in a high degree of
reproducibility of the measurements. Depending on the ionic
concentration as well as the amplitude of the applied voltage, the
observed direction of the DC flow component is either forward or
reverse. The impedance spectrum has been thoroughly measured and
analyzed in terms of an equivalent circuit diagram. Our
observations agree qualitatively, but not quantitatively, with
theoretical models published in the literature.
\end{abstract}

\maketitle

\section{Introduction}\label{sec:introduction}
The recent interest in AC electrokinetic micropumps was initiated
by experimental observations by Green, Gonzales \emph{et al.} of
fluid motion induced by AC electroosmosis over pairs of
microelectrodes \cite{Green2000,Gonzales2000,Green2002} and by a
theoretical prediction by Ajdari that the same mechanism would
generate flow above an electrode array \cite{Ajdari2000}. Brown
\emph{et al.}~\cite{Brown2000} demonstrated experimentally pumping
of electrolyte with a low voltage, AC biased electrode array, and
soon after the same effect was reported by a number of other
groups observing flow velocities of the order of mm/s
\cite{Studer2002,Mpholo2003,Lastochkin2004,Debesset2004,Studer2004,Cahill2004,Ramos2005,Garcia2006}.
Several theoretical models have been proposed parallel to the
experimental observations \cite{Ramos2003,Mortensen2005,LHO2006}.
However, so far not all aspects of the flow-generating mechanisms
have been explained.

Studer \emph{et al.}~\cite{Studer2004} made a thorough investigation
of flow dependence on electrolyte concentration, driving voltage and
frequency for a characteristic system. In this work a reversal of
the pumping direction for frequencies above 10~kHz and rms voltages
above 2~V was reported. For a travelling wave device Ramos \emph{et
al.}~\cite{Ramos2005} observed reversal of the pumping direction at
1~kHz and voltages above 2~V. The reason for this reversal is not
yet fully understood and the goal of this work is to contribute with
further experimental observations of reversing flow for other
parameters than those reported previously.

An integrated electrokinetic AC driven micropump has been fabricated
and studied. The design follows Studer \emph{et
al.}~\cite{Studer2004}, where an asymmetric array of electrodes
covers the channel bottom in one section of a closed pumping loop.
Pumping velocities are measured in another section of the channel
without electrodes. In this way electrophoretic interaction between
the beads used as flow markers and the electrodes is avoided. In
contrast to the soft lithography utilized by Studer \emph{et al.},
we use more well-defined MEMS fabrication techniques in Pyrex glass.
This results in a very robust system, which exhibits stable
properties and remains functional over time periods extending up to
a year. Furthermore, we have a larger electrode coverage of the
total channel length allowing for the detection of smaller pumping
velocities. Our improved design has led to the observation of a new
phenomenon, namely the reversing of the flow at low voltages and low
frequencies. The electrical properties of the fabricated
microfluidic chip have been investigated to clarify whether these
reflect the reversal of the flow direction. In accordance with the
electrical measurements we propose and evaluate an equivalent
circuit diagram. Supplementary details related to the present work
can be found in Ref.~\cite{MIGthesis}.

\section{Experimental}\label{sec:experimental}
\subsection{System design}
\begin{figure}[!t]
\centering
\includegraphics[width=\columnwidth]{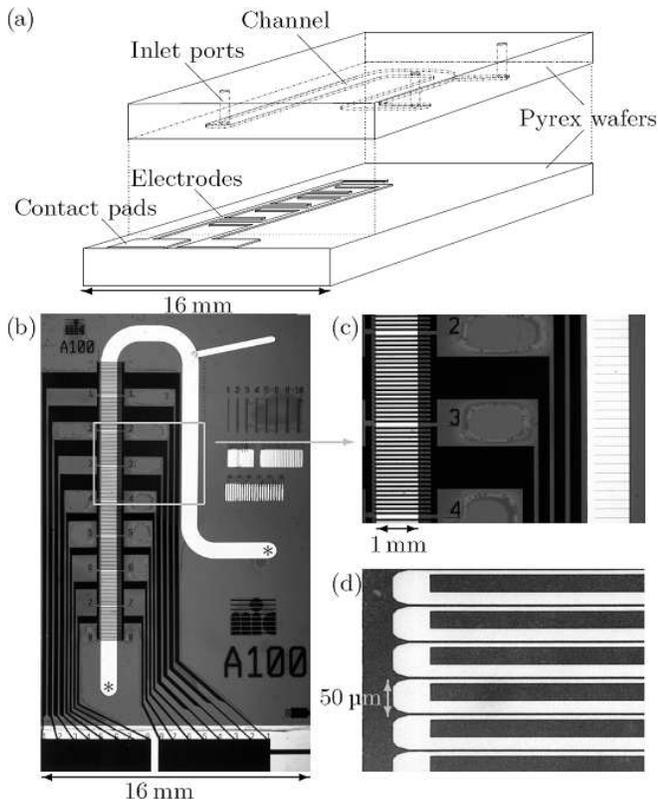}
\caption{(a) Sketch of the fabricated chip consisting of two Pyrex
glass wafers bonded together. The channels are etched into the top
wafer, which also contains the fluid access ports. Flow-generating
electrodes are defined on the bottom wafer. (b) Micrograph of the
full chip containing a channel (white) with flow-generating
electrodes (black) and a narrow side channel for bead injection
(upper right corner). During flow measurements the channel ends
marked with an asterisk are connected by an outer tube. The
electrode array is divided into eight sub arrays, each having its
own connection to the electrical contact pad. (c) Magnification of
the framed area in panel (b) showing the flow-generating electrodes
to the left and the measurement channel with ruler lines to the
right. (d) Close up of an electrode array section with electrode
translation period of 50~$\mu$m.} \label{fig:ChipA100}
\end{figure}

The microchip was fabricated for studies of the basic electrokinetic
properties of the system. Hence, a simple microfluidic circuit was
designed to eliminate potential side-effects due to complex device
issues. The chip consists of two 500~$\mu$m thick Pyrex glass wafers
anodically bonded together. Metal electrodes are defined on the
bottom wafer and channels are contained in the top wafer, as
illustrated schematically in Fig.~\ref{fig:ChipA100}(a). This
construction ensures an electrical insulated chip with fully
transparent channels.

An electrode geometry akin to the one utilized by Brown \emph{et
al.} \cite{Brown2000} and Studer \emph{et al.}~\cite{Studer2004} was
chosen.  The translation period of the electrode array is 50~$\mu$m
with electrode widths of $W_1= 4.2~\mu$m and $W_2= 25.7~\mu$m, and
corresponding electrode spacings of $G_1= 4.5~\mu$m and $G_2=
15.6~\mu$m, see Fig.~\ref{fig:ChipA100}(d). Further theoretical
investigations have shown that this geometry results in a nearly
optimal flow velocity \cite{LHO2006}. The total electrode array
consists of eight sub-arrays each having their own connection to the
shared contact pad, Fig.~\ref{fig:ChipA100}(b). This construction
makes it possible to disconnect a malfunctioning sub-array. The
entire electrode array has a width of 1.3~mm ensuring that the
alignment of the electrodes and the 1.0~mm wide fluidic channels is
not critical.

A narrow side channel, Fig.~\ref{fig:ChipA100}(b), allows beads to
be introduced into the part of the channel without electrodes, where
a number of ruler lines with a spacing of 200~$\mu$m enable flow
measurements by particle tracing, Fig.~\ref{fig:ChipA100}(c).

An outer circuit of valves and tubes is utilized to control and
direct electrolytes and bead solutions through the channels. During
flow-velocity measurements, the inlet to the narrow side channel is
blocked and to eliminate hydrostatic pressure differences the two
ends of the main channel are connected by an outer teflon tube with
an inner diameter of 0.5~mm. The hydraulic resistance of this outer
part of the pump loop is three orders of magnitude smaller than the
on-chip channel resistance and is thus negligible.

The maximal velocity of the Poiseuille flow in the measurement
channel section is denoted $v_\textrm{pois}$, and the average slip
velocity generated above the electrodes by electroosmosis is denoted
$v_\textrm{slip}$. To obtain a measurable $v_\textrm{pois}$ at as
low applied voltages as possible, the electrode coverage of the
total channel length is made as large as possible. In our system the
total channel length is $L_\textrm{tot}= 40.8~\textrm{mm}$ and the
section containing electrodes is $L_\textrm{el}= 16.0~\textrm{mm}$,
which ensures a high Poiseuille flow velocity, $v_\textrm{pois} =
(3/4)(L_\textrm{el}/L_\textrm{tot})v_\textrm{slip} =
0.29\,v_\textrm{slip}$ \cite{MIGthesis}.

The microfluidic chip has a size of approximately
$16~\textrm{mm}\times 28~\textrm{mm}$ and is shown in
Fig.~\ref{fig:ChipA100}, and the device parameters are listed in
Table~\ref{tab:geometry}.

\begin{table}[!t]
\centering
\begin{ruledtabular}
\begin{tabular}{l*{2}{>{$}c<{$}}}
 Channel height                           &        H         &        33.6~\mu\textrm{m}         \\
 Channel width                            &       w          &         967~\mu\textrm{m}         \\
 Channel length                           &  L_\textrm{tot}  &         40.8~\textrm{mm}              \\
 Channel length with electrodes           &  L_\textrm{el}   &         16.0~\textrm{mm} \vspace{2ex} \\
 Width of electrode array                 &   w_\textrm{el}  &         1300~\mu\textrm{m}        \\
 Narrow electrode gap                     &       G_1        &         4.5~\mu\textrm{m}         \\
 Wide electrode gap                       &       G_2        &        15.6~\mu\textrm{m}         \\
 Narrow electrode width                   &       W_1        &         4.2~\mu\textrm{m}         \\
 Wide electrode width                     &       W_2        &        25.7~\mu\textrm{m}         \\
 Electrode thickness                      &        h         &        0.40~\mu\textrm{m}         \\
 Electrode surface area ($[W_1+2h]w$)     &     A_1          &   4.84\times 10^{-9}~\textrm{m}^2                \\
 Electrode surface area ($[W_2+2h]w$)     &     A_2          &   25.63\times 10^{-9}~\textrm{m}^2               \\
 Number of electrode pairs                &        p         &                312                 \\
 Electrode resistivity (Pt)               &       \rho       & 10.6\times 10^{-8}~\Omega\textrm{m} \vspace{2ex} \\
 Electrolyte conductivity (0.1~mM) &      \sigma      &          1.43~\textrm{mS/m}           \\
 Electrolyte conductivity (1.0~mM) &      \sigma      &          13.5~\textrm{mS/m}           \\
 Electrolyte permittivity                 &    \epsilon      &          80\,\epsilon_0            \\
 Pyrex permittivity                       & \epsilon_ \textrm{p} &      4.6\,\epsilon_0           \\
\end{tabular}
\end{ruledtabular} \caption{Dimensions and parameters of the
fabricated microfluidic system.}\label{tab:geometry}
\end{table}

\subsection{Chip fabrication}
The flow-generating electrodes of e-beam evaporated
Ti(10~nm)/Pt(400~nm) were defined by lift-off in 1.5~$\mu$m thick
photoresist AZ 5214-E (Hoechst) using a negative process. The Ti
layer ensures good adhesion to the Pyrex substrate. Platinum is
electrochemically stable and has a low resistivity, which makes it
suitable for the application. By choosing an electrode thickness of
$h= 400$~nm, the metallic resistance between the contact pads and
the channel electrolyte is at least one order of magnitude smaller
than the resistance of the bulk electrolyte covering the electrode
array.

In the top Pyrex wafer the channel of width $w= 967~\mu$m and height
$H= 33.6~\mu$m was etched into the surface using a solution of 40\%
hydrofluoric acid. A 100~nm thick amorphous silicon layer was
sputtered onto the wafer surface and used as etch mask in
combination with a 2.2~$\mu$m thick photoresist layer. The channel
pattern was defined by a photolithography process akin to the
process used for electrode definition, and the wafer backside and
edges were protected with a 70~$\mu$m thick etch resistant PVC foil.
The silicon layer was then etched away in the channel pattern using
a mixture of nitric acid and buffered hydrofluoric acid,
HNO$_3$:BHF:H$_2$O = 20:1:20. The wafer was subsequently baked at
120$^\circ$C to harden the photoresist prior to the HF etching of
the channels. Since the glass etching is isotropic, the channel
edges were left with a rounded shape. However, this has only a minor
impact on the flow profile, given that the channel aspect ratio is
$w/H \approx 30$. The finished wafer was first cleaned in acetone,
which removes both the photoresist and the PVC foil, and then in a
piranha solution.

After alignment of the channel and the electrode array, the two chip
layers were anodically bonded together by heating the ensemble to
400$^\circ$C and applying a voltage difference of 700~V across the
two wafers for 10~min. During this bonding process, the previously
deposited amorphous Si layer served as diffusion barrier against the
sodium ions in the Pyrex glass. Finally, immersing the chip in
DI-water holes were drilled for the in- and outlet ports using a
cylindrical diamond drill with a diameter of 0.8~mm.

\subsection{Measurement setup and procedures}
\begin{figure}[!b]
\centering
\includegraphics[scale=1]{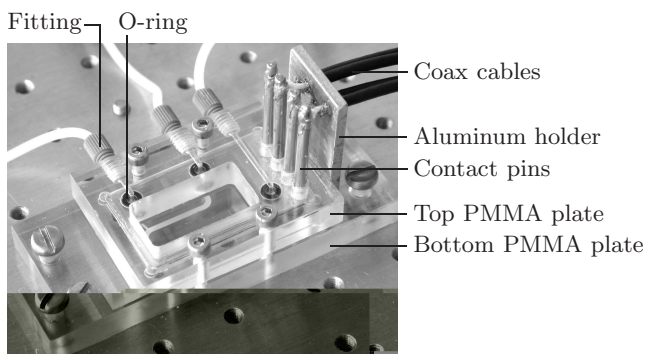}
\caption{Chip holder constructed to connect external tubing and
electrical wiring with the microfluidic chip.}
\label{fig:ChipHolder}
\end{figure}

Liquid injection and electrical contact to the microchip was
established through a specially constructed PMMA chip holder, shown
in Fig.~\ref{fig:ChipHolder}. Teflon tubing was fitted into the
holder in which drilled channels provided a connection to the
on-chip channel inlets. The interface from the chip holder to the
chip inlets was sealed by O-rings. Electrical contact was obtained
with spring loaded contact pins fastened in the chip holder and
pressed against the electrode pads. The inner wires of thin coax
cables were soldered onto the pins and likewise fastened to the
holder.

The pumping was induced by electrolytic solutions of KCl in
concentrations ranging from $c= 0.1$~mM to 1.0~mM. The chip was
prepared for an experiment by careful injection of this electrolyte
into the channel and tubing system, after which the three valves to
in- and outlets were closed. The electrical impedance spectrum of
the microchip was measured before and after each series of flow
measurements to verify that no electrode damaging had occurred
during the experiments. If the impedance spectrum had changed, the
chip and the series of performed measurements were discarded.
Velocity measurements were only carried out when the tracer beads
were completely at rest before biasing the chip, and it was always
verified that the beads stopped moving immediately after switching
off the bias. The steady flow was measured for 10~s to 30~s. After a
series of measurements was completed, the system was flushed
thoroughly with milli-Q water. When stored in milli-Q water between
experiments the chips remained functional for at least one year.

\subsection{AC biasing and impedance measurements}
Using an impedance analyzer (HP 4194 A), electrical impedance
spectra of the microfluidic chip were obtained by four-point
measurements, where each contact pad was probed with two contact
pins. Data was acquired from 100~Hz to 15~MHz. To avoid electrode
damaging by application of a too high voltage at low frequencies,
all impedance spectra were measured at $V_\textrm{rms}= 10$~mV.

The internal sinusoidal output signal of a lock-in amplifier
(Stanford Research SR830DSP) was used for AC biasing of the
electrode array during flow-velocity measurements. The applied rms
voltages were in the range from 0.5~V to 2~V and the frequencies
between 0.5~kHz and 100~kHz. A current amplification was necessary
to maintain the correct potential difference across the electrode
array, since the overall chip resistance could be small ($\sim
0.1~\textrm{k}\Omega$ to 1~k$\Omega$) when frequencies in the given
interval were applied. The current through the microfluidic chip was
measured by feeding the output signal across a small series resistor
back into the lock-in amplifier.

The lock-in amplifier was also used for measuring impedance spectra
for frequencies below 100~Hz, which were beyond the span of the
impedance analyzer.

\subsection{Flow velocity measurements}
After filling the channel with an electrolyte and actuating the
electrodes, the flow measurements were performed by tracing beads
suspended in the electrolyte.

Fluorescent beads (Molecular Probes, FluoSpheres F-8765) with a
diameter of 1~$\mu$m were introduced into the measurement section of
the channel and used as flow markers for the velocity determination.
A stereo microscope was focused at the beads, and with an attached
camera pictures were acquired with time intervals of $\Delta t =
0.125$~s to 1.00~s depending on the bead velocity. Subsequently, the
velocity was determined by averaging over a distance of $\Delta x =
200~\mu$m, i.e., $v=\Delta x/\Delta t$. Only the fastest beads were
used for flow detection, since these are assumed to be located in
the vertical center of the channel. It should be noted that the use
of fluorescent particles prevented an introduction of significant
illumination heating of the sample.

The limited number of acquired pictures led to an uncertainty of 5\%
in the determination of flow velocities, which corresponds to the
movement of the tracer beads within $0.5$ to $1$ frame.
Additionally, there is a statistical uncertainty on the vertical
particle position in the channel, which is estimated to introduce up
to 10\% error on the determined bead velocity. It is then assumed
that the fastest beads are positioned within $H/3$ of the maximum of
the Poiseuille flow profile.

\section{Results}\label{sec:results}
\begin{figure}[!t]
\centering
\includegraphics[scale=1]{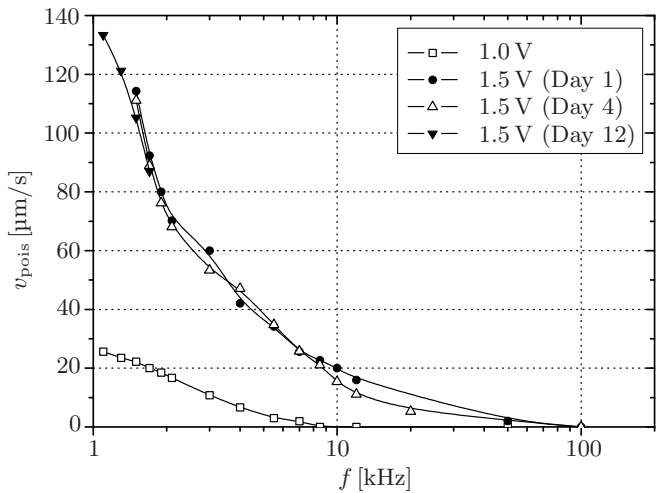}
\caption{Reproducible flow-velocities induced in a 0.1~mM KCl
solution and observed at different days as a function of frequency
at a fixed rms voltage of 1.5~V. A corresponding series was measured
at $V_\textrm{rms}= 1.0$~V. Lines have been added to guide the eye.}
\label{fig:Reproducibility}
\end{figure}

In the parameter ranges corresponding to those published in the
literature, our flow velocity measurements are in agreement with
previously reported results. Using a $c= 0.1$~mM KCl solution and
driving voltages of $V_\textrm{rms}= 1.0$~V to 1.5~V over a
frequency range of $f= 1.1$~kHz to 100~kHz, we observed among other
measurement series the pumping velocities shown in
Fig.~\ref{fig:Reproducibility}. The general tendencies were an
increase of velocity towards lower frequencies and higher voltages,
and absence of flow above $f\sim 100$~kHz. The measured velocities
corresponded to slightly more than twice those measured by Studer
\emph{et al.}~\cite{Studer2004} due to our larger electrode coverage
of the total channel. We observed damaging of the electrodes if more
than 1~V was applied to the chip at a driving frequency below 1~kHz,
for which reason there are no measurements at these frequencies. It
is, however, plausible that the flow velocity for our chip peaked
just below $f\sim 1$~kHz.

\begin{figure}[!t]
\centering
\includegraphics[scale=1]{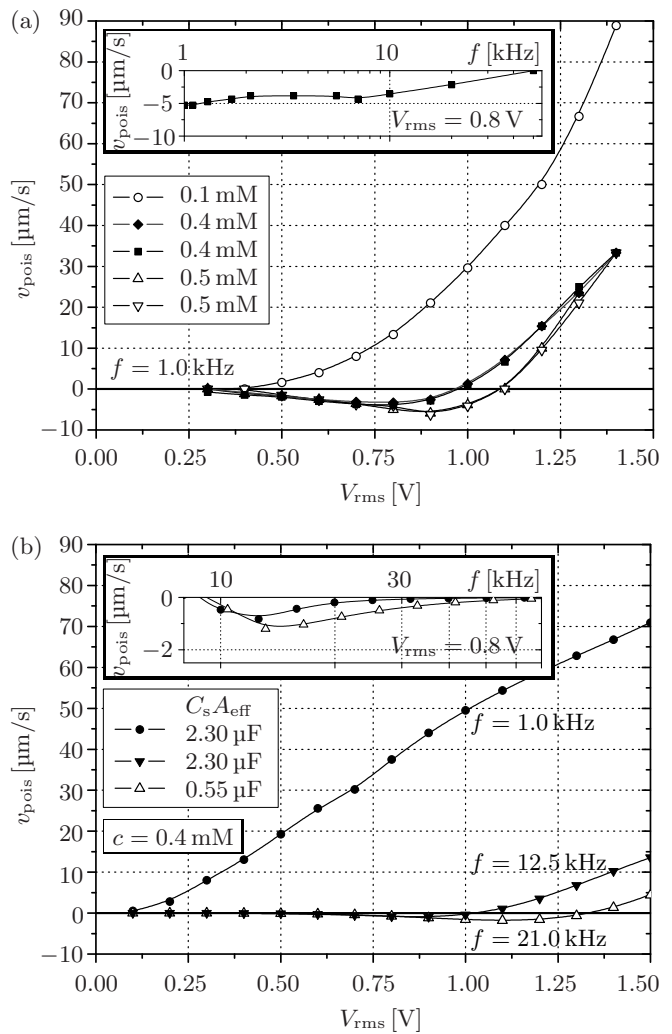}
\caption{(a) Reversed flow observed for repeated measurements of two
concentrations of KCl at 1.0~kHz. The inset shows that for a 0.4~mM
KCl solution at a fixed rms voltage of 0.8~V the flow direction
remains negative, but slowly approaches zero for frequencies up to
50~kHz. (b) The theoretical model presented in Ref.~\cite{LHOthesis}
predicts the trends of the experimentally observed velocity curves.
The depicted graphs are calculated for a $c= 0.4$~mM solution and
parameters corresponding to the experiments
(Table~\ref{tab:equivalent}) with $\zeta_\textrm{eq}= 160$~mV.
Additional curves have been plotted for slightly different parameter
values in order to obtain a closer resemblance to the experimental
graphs, see Sec.~\ref{sec:discussion}. }\label{fig:Reverseflow}
\end{figure}

\subsection{Reproducibility of measurements}
Our measured flow velocities were very reproducible due to the
employed MEMS chip fabrication techniques and the careful
measurement procedures described in Sec.~\ref{sec:experimental}.
This is illustrated in Fig.~\ref{fig:Reproducibility}, which shows
three velocity series recorded several days apart. The measurements
were performed on the same chip and for the same parameter values.
Between each series of measurements, the chip was dismounted and
other experiments performed. However, it should be noted that a very
slow electrode degradation was observed when a dozen of measurement
series were performed on the same chip over a couple of weeks.

\subsection{Low frequency reversed flow}
Devoting special attention to the low-frequency ($f< 20$~kHz),
low-voltage regime ($V_\textrm{rms}< 2$~V), not studied in detail
previously, we observed an unanticipated flow reversal for certain
parameter combinations. Fig.~\ref{fig:Reverseflow}(a) shows flow
velocities measured for a frequency of 1.0~kHz as a function of
applied voltage for various electrolyte concentrations. It is
clearly seen that the velocity series of $c= 0.1$~mM exhibits the
known exclusively forward and increasing pumping velocity as
function of voltage, whereas for slightly increased electrolyte
concentrations an unambiguous reversal of the flow direction is
observed for rms voltages below approximately 1~V.

This reversed flow direction was observed for all frequencies in the
investigated spectrum when the electrolyte concentration and the rms
voltage were kept constant. This is shown in the inset of
Fig.~\ref{fig:Reverseflow}(a), where a velocity series was obtained
over the frequency spectrum for an electrolyte concentration of
0.4~mM at a constant rms voltage of 0.8~V. It is noted that the
velocity is nearly constant over the entire frequency range and
tends to zero above $f\sim 20$~kHz.

\begin{figure}[!t]
\centering
\includegraphics[scale=1]{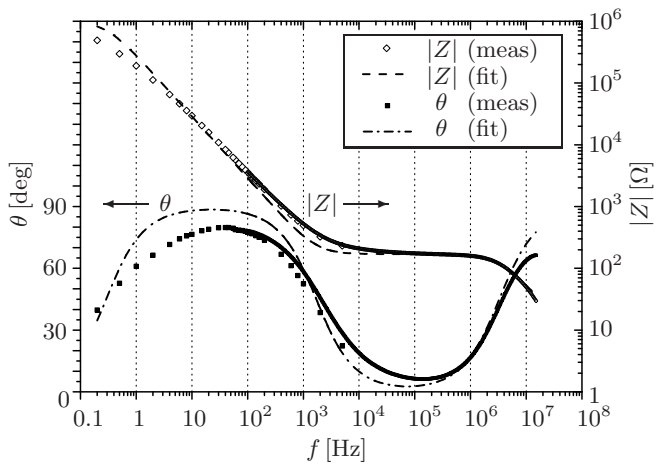}
\caption{Bode plot showing the measured amplitude $|Z|$ (right
ordinate axis) and phase $\theta$ (left ordinate axis) of the
impedance as a function of frequency over eight decades from 0.2~Hz
to 15~MHz. The voltage was $V_\textrm{rms}= 10$~mV and the
electrolyte concentration $c= 1.0$~mM KCl. The measurements are
shown with symbols while the curves of the fitted equivalent diagram
are represented by dashed lines. The measurement series obtained
with the Impedance Analyzer consist of 400 very dense points while
the series measured using the lock-in amplifier contains fewer
points with a clear spacing. }\label{fig:Bode}
\end{figure}

\subsection{Electrical characterization}
To investigate whether the flow reversal was connected to unusual
properties of the electrical circuit, we carefully measured the
impedance spectrum $Z(f)$ of the microfluidic system. Spectra were
obtained for the chip containing KCl electrolytes with the different
concentrations $c= 0.1$~mM, 0.4~mM and 1.0~mM.

Fig.~\ref{fig:Bode} shows the Bode plots of the impedance spectrum
obtained for $c= 1.0$~mM. For frequencies between $f\sim 1$~Hz and
$f\sim 10^3$~Hz the curve shape of the impedance amplitude $|Z|$ is
linear with slope $-1$, after which a horizontal curve section
follows, and finally the slope again becomes $-1$ for frequencies
above $f\sim 10^6$~Hz. Correspondingly, the phase $\theta$ changes
between $0^{\circ}$ and $90^{\circ}$. From the decrease in phase
towards low frequencies it is apparent that $|Z|$ must have another
horizontal curve section below $f\sim 1$~Hz. When the curve is
horizontal and the phase is $0^{\circ}$ the system behaves
resistively, while it is capacitively dominated when the phase is
$90^{\circ}$ and the curve has a slope of $-1$.

\begin{figure}[!t]
\centering
\includegraphics[width=\linewidth]{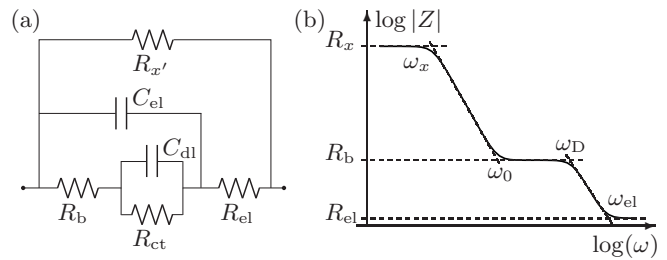}
\caption{(a) Equivalent circuit diagram. (b) Sketch of the impedance
amplitude curve of the equivalent diagram. It consists of three
plateaus, and four characteristic frequencies $\omega_x$,
$\omega_0$, $\omega_\textrm{D}$ and $\omega_\textrm{el}$ (see
Table~\ref{tab:symbols}) that characterize the shape and may be
utilized to estimate the component values.}
\label{fig:EquivalentDiagram}
\end{figure}

\begin{table}[!b]
\centering
\begin{ruledtabular}
\begin{tabular}{l l}
 Debye length & $\lambda_\textrm{D}$\\
  Total electrode resistance & $R_\textrm{el}$\\
 Total bulk electrolyte resistance & $R_\textrm{b}$\\
 Total faradaic (charge transfer) resistance & $R_\textrm{ct}$\\
 Internal resistance in lock-in amplifier & $R_{x'}$\\
 Total measured resistance for $\omega\to 0$ & $R_x$\\
 Total electrode capacitance & $C_\textrm{el}$\\
 Total double layer capacitance & $C_\textrm{dl}$\\
 Debye layer capacitance & $C_\textrm{D}$\\
 Surface capacitance & $C_\textrm{s}$\vspace{2ex} \\
 Debye frequency & $\omega_\textrm{D}$\\
 Inverse ohmic relaxation time  & $\omega_0$\\
 Inverse faradaic charge transfer time (primarily)  & $\omega_x$\\
 Characteristic frequency of electrode circuit & $\omega_\textrm{el}$\\
\end{tabular}
\end{ruledtabular} \caption{List of the symbols used in the
equivalent circuit model.}\label{tab:symbols}
\end{table}

\begin{table*}[!t]
\newcolumntype{Z}{>{\centering $}p{0.65cm}<{$}}
\centering
\begin{ruledtabular}
\begin{tabular}{>{$}p{2cm}<{$} ZZ@{\hspace{0.6cm}}ZZ@{\hspace{0.6cm}}Z@{\hspace{0.6cm}}ZZ@{\hspace{0.6cm}}ZZ@{\hspace{0.6cm}}ZZ@{\hspace{0.6cm}}ZZ}
                             & R_\textrm{b}   & R_\textrm{b}   & R_\textrm{el}  & R_\textrm{el}  & R_\textrm{ct}  & C_\textrm{dl}  & C_\textrm{dl}  & C_\textrm{el}  & C_\textrm{el}  & \omega_\textrm{D} & \omega_\textrm{D} & \omega_\textrm{0} & \omega_\textrm{0}  \tabularnewline [-1.5mm]
                             & _\textrm{mod}  & _\textrm{meas} & _\textrm{mod}  & _\textrm{meas} & _\textrm{meas} & _\textrm{mod}  & _\textrm{meas} & _\textrm{mod}  & _\textrm{meas} & _\textrm{mod}     & _\textrm{meas}    & _\textrm{mod}     & _\textrm{meas}     \tabularnewline
                             &[\textrm{k}\Omega]&[\textrm{k}\Omega]& [\Omega]   & [\Omega]       &[\textrm{M}\Omega]&[\mu\textrm{F}]&[\mu\textrm{F}]& [\textrm{nF}]  & [\textrm{nF}]  &\multicolumn{2}{l}{[M\,rad\,s$^{-1}$]}&\multicolumn{2}{c}{[k\,rad\,s$^{-1}$]} \vspace{0.5ex} \tabularnewline
\hline
 \rule{0pt}{3ex}
 0.1~\textrm{mM}\,(\textrm{A})  &    2.0         &    1.0         &     7.6        &     5          &    1.0         &    0.50        &   0.50         &    0.28        &    0.30        &      2.0          &      3.3          &      1.0          &      2.0           \tabularnewline
 1.0~\textrm{mM}\,(\textrm{A})  &    0.21        &    0.17        &     7.6        &     6          &    1.0         &    0.56        &   0.55         &    0.28        &    0.29        &      19.1         &      20.6         &      8.5          &      10.7     \vspace{2ex} \tabularnewline
 0.1~\textrm{mM}\,(\textrm{B})  &    2.0         &    1.4         &     7.6        &     6          &      -         &    0.50        &   0.51         &    0.28        &    0.29        &      2.0          &      3.0          &      1.0          &      1.4           \tabularnewline
 0.4~\textrm{mM}\,(\textrm{B})  &    0.52        &    0.41        &     7.6        &     7          &      -         &    0.54        &   0.53         &    0.28        &    0.28        &      7.7          &      9.3          &      3.6          &      4.6           \tabularnewline
 1.0~\textrm{mM}\,(\textrm{B})  &    0.21        &    0.17        &     7.6        &     8          &      -         &    0.56        &   0.55         &    0.28        &    0.26        &      19.1         &      22.6         &      8.5          &      10.5          \tabularnewline
\end{tabular}
\end{ruledtabular}
\caption{Comparison of measured (meas) and modeled (mod) values of
the components in the equivalent diagram,
Fig.~\ref{fig:EquivalentDiagram}. The measured values are given by
curve fits of Bode plots, Fig.~\ref{fig:Bode}, obtained on two
similar chips labeled A and B, respectively. The modeled values are
estimated on basis of Table~\ref{tab:geometry} and a particular
choice of the parameters $\zeta_\textrm{eq}$ and $C_\textrm{s}$.
This choice is not unique since different combinations can lead to
the same value of $C_\textrm{dl}$. }\label{tab:equivalent}
\end{table*}

\subsection{Equivalent circuit}
In electrochemistry the standard way of analyzing such impedance
measurements is in terms of an equivalent circuit diagram
\cite{BardFaulkner}. The choice of diagram is not unambiguous
\cite{Green2002}. We have chosen the diagram shown in
Fig.~\ref{fig:EquivalentDiagram}(a) with the component labeling
listed in Table~\ref{tab:symbols}.

Charge transport through the bulk electrolyte is represented by an
ohmic resistance $R_\textrm{b}$, accumulation of charge in the
double layer at the electrodes by a capacitance $C_\textrm{dl}$, and
faradaic current injection from electrochemical reactions at the
electrodes by another resistance, the charge-transfer resistance
$R_\textrm{ct}$ \cite{BardFaulkner,LHO2006}. Moreover, we include
the ohmic resistance of the metal electrodes $R_\textrm{el}$, the
mutual capacitance between the narrow and wide electrodes
$C_\textrm{el}$, and a shunt resistance $R_{x'} = 10$~M$\Omega$ to
represent the internal resistance of the lock-in amplifier.

Finally, in electrochemical experiments at low frequency, the
electrical current is often limited by diffusive transport of the
reactants in the faradaic electrode reaction to and from the
electrodes. This can be modeled by adding a frequency dependent
Warburg impedance in series with the charge transfer resistance
\cite{BardFaulkner}. However, because the separation between the
electrodes is so small and the charge transfer resistance is so
large, we are unable to distinguish the Warburg impedance in the
impedance measurements and leave it out of the equivalent diagram.

By fitting the circuit model to the impedance measurements we
extract the component values listed in Table~\ref{tab:equivalent}.
On the chip labeled B we were unable to measure the charge transfer
resistance due to a minor error on the chip introduced during the
bonding process. Fig.~\ref{fig:EquivalentDiagram}(b) illustrates the
relation between component values and the impedance amplitude curve
through four characteristic angular frequencies $\omega = 2\pi f$.
The inverse frequency $\omega_x^{-1}= R_x\,C_\textrm{dl}$ primarily
expresses the characteristic time for the faradaic charge transfer
into the Debye layer. The characteristic time for charging the Debye
layer through the electrolyte is given by
$\omega_0^{-1}=R_\textrm{b}\,C_\textrm{dl}$. The Debye frequency is
$\omega_\textrm{D}=1/(R_\textrm{b}\,C_\textrm{el})$, and finally
$\omega_\textrm{el}=1/(R_\textrm{el}\,C_\textrm{el})$ simply states
the characteristic frequency for the on-chip electrode circuit in
the absence of electrolyte. It is noted that the total DC-limit
resistance $R_x$ corresponds to the parallel coupling between
$R_{x'}$ and $R_\textrm{ct}$.

\section{Discussion}\label{sec:discussion}
In the following we investigate to which extent the general theory
of induced-charge (AC) electroosmosis can explain our observations
and experimental data. We first use the equivalent circuit component
values extracted from the impedance measurements to estimate some
important electrokinetic parameters based on the
Gouy--Chapman--Stern model \cite{BardFaulkner}, namely, the Stern
layer capacitance $C_\textrm{s}$, the intrinsic zeta potential
$\zeta_\textrm{eq}$ on the electrodes and the charge transfer
resistance $R_\textrm{ct}$. Then we use this as input to the weakly
nonlinear electro-hydrodynamic model presented in
Ref.~\cite{LHOthesis}, which is an extension of the model in
Ref.~\cite{LHO2006}. We compare theoretical values with experimental
observations, and discuss the experimentally observed trends of the
flow velocities.

\subsection{Impedance analysis}
The impedance measurements are performed at a low voltage of
$V_\textrm{rms}= 10$~mV so it might be expected that Debye--Hückel
linear theory applies ($V\lesssim 25$~mV). However, since we only
measure the potential difference between the electrodes and we do
not know the potential of the bulk electrolyte, we cannot say much
about the exact potential drop across the double layer. Many
electrode-electrolyte systems posses an intrinsic zeta potential at
equilibrium $\zeta_\textrm{eq}$ of up to a few hundred mV. Indeed,
the measured $C_\textrm{dl}$ is roughly $10$ times larger than
predicted by Debye--Hückel theory, which indicates that the
intrinsic zeta potential is at least $\pm 125$~mV.

According to Gouy--Chapman--Stern theory the $C_\textrm{dl}$ can be
expressed as a series coupling of the compact Stern layer
capacitance $C_\textrm{s}$ and the differential Debye-layer
capacitance $C_\textrm{D}$,
\begin{equation}\label{eq:Cdl}
\frac{A_\textrm{eff}}{C_\textrm{dl}} = \frac{1}{C_\textrm{s}} +
\frac{1}{C_\textrm{D}},
\end{equation}
where the two double-layer capacitances of an electrode pair are
coupled in series through the electrolyte, and since the $p$
electrode pairs are coupled in parallel, the effective area of the
total double layer is $A_\textrm{eff}=p\,A_1 A_2 /(A_1+A_2)$. $A_1$
and $A_2$ are the total surface areas exposed to the electrolyte of
a narrow and wide electrode, respectively. For simplicity
$C_\textrm{s}$ is often assumed constant and independent of
potential and concentration, while $C_\textrm{D}$ is given by the
Gouy--Chapman theory as
\begin{equation}
C_\textrm{D}=\frac{\epsilon}{\lambda_\textrm{D}}
\cosh\left(\frac{\zeta_\textrm{eq} z e}{ 2 k_\textrm{B} T}\right).
\end{equation}
Unfortunately, it is not possible to estimate the exact values of
both $C_\textrm{s}$ and $\zeta_\textrm{eq}$ from a measurement of
$C_\textrm{dl}$, because a range of parameters lead to the same
$C_\textrm{dl}$. We can, nevertheless, state lower limits as
$C_\textrm{s}\geq 0.39$~F/m$^2$ and $|\zeta_\textrm{eq}|\geq 175$~mV
for $c= 0.1$~mM or $C_\textrm{s}\geq 0.43$~F/m$^2$ and
$|\zeta_\textrm{eq}|\geq 125$~mV at $c= 1.0$~mM.

For the model values in Table~\ref{tab:equivalent} we used
Eq.~(\ref{eq:Cdl}) with $C_\textrm{s} = 1.8$~F/m$^2$ and
$\zeta_\textrm{eq}= 190$~mV, 160~mV and 140~mV at 0.1~mM, 0.4~mM and
1.0~mM KCl, respectively, in accordance with the trend often
observed that $\zeta_\textrm{eq}$ decreases with increasing
concentration, \cite{Kirby2004}. The bulk electrolyte resistance can
be expressed as
\begin{equation}
 R_\textrm{b}=\frac{0.85}{\sigma w p},
\end{equation}
where $\sigma$ is the conductivity, $w$ is the width of the
electrodes and $p$ is the number of electrode pairs, see
Table~\ref{tab:geometry}, and $0.85$ is a numerical factor computed
for our particular electrode layout using the finite-element based
program \textsc{Comsol Multiphysics}. Similarly, the mutual
capacitance between the electrodes can be calculated as
\begin{equation}
 C_\textrm{el}=\frac{p}{0.85}\big[\,\epsilon w +
 \epsilon_\textrm{p}(2w_\textrm{el}-w)\big],
\end{equation}
and the resistance $R_\textrm{el}$ of the electrodes leading from
the contact pads to the array is simply estimated from the
resistivity of platinum and the electrode geometry.

At frequencies above 100~kHz the impedance is dominated by
$R_\textrm{b}$, $C_\textrm{el}$ and $R_\textrm{el}$, and the Bode
plot closely resembles a circuit with ideal components, see
Fig.~\ref{fig:Bode}. Around 1~kHz we observe some frequency
dispersion which could be due to the change in electric field line
pattern around the inverse RC-time $\omega_0 =
1/(R_\textrm{b}C_\textrm{dl})$ \cite{LHOthesis}. Finally, below
1~kHz where the impedance is dominated by $C_\textrm{dl}$, the phase
never reaches $90^{\circ}$ indicating that the double layer
capacitance does not behave as an ideal capacitor but more like a
constant phase element (CPE). This behavior is well known
experimentally, but not fully understood theoretically
\cite{Kerner2000}.

\subsection{Flow}
The forward flow velocities measured at $c= 0.1$~mM as a function of
frequency, Fig.~\ref{fig:Reproducibility}, qualitatively exhibit the
trends predicted by standard theory, namely, the pumping increases
with voltage and falls off at high frequency
\cite{Ajdari2000,Ramos2003}.

More specifically, the theory predicts that the pumping velocity
should peak at a frequency around the inverse RC-time $\omega_0$,
corresponding to $f\approx 0.3$~kHz, and decay as the inverse of the
frequency for our applied driving voltages, see Fig.~11 in
Ref.~\cite{LHO2006}. Furthermore, the velocity is predicted to grow
like the square of the driving voltage at low voltages, changing to
$V\log V$ at large voltages \cite{LHO2006,LHOthesis}.

Experimentally, the velocity is indeed proportional to $\omega^{-1}$
and the peak is not observed within the range 1.1~kHz to 100~kHz,
but it is likely to be just below 1~kHz. However, the increase in
velocity between 1.0~V and 1.5~V displayed in
Fig.~\ref{fig:Reproducibility} is much faster than $V^2$. That is
also the result in Fig.~\ref{fig:Reverseflow}(a) for $c= 0.1$~mM
where no flow is observed below $V_\textrm{rms}= 0.5$~V, while above
that voltage the velocity increases rapidly. For $c= 0.4$~mM and $c=
0.5$~mM the velocity even becomes negative at voltages
$V_\textrm{rms}\leq 1$~V. This cannot be explained by the standard
theory and is also rather different from the reverse flow that has
been observed by other groups at larger voltages $V_\textrm{rms}>
2$~V and at frequencies above the inverse RC-time
\cite{Studer2004,Ramos2005,Garcia2006}.

The velocity shown in the inset of Fig.~\ref{fig:Reverseflow}(a) is
remarkable because it is almost constant between 1~kHz and 10~kHz.
This is unlike the usual behavior for AC electroosmosis that always
peaks around the inverse RC-time, because it depends on partial
screening at the electrodes to simultaneously get charge and
tangential field in the Debye layer. At lower frequency the
screening is almost complete so there is no electric field in the
electrolyte to drive the electroosmotic fluid motion, while at
higher frequency the screening is negligible so there is no charge
in the Debye layer and again no electroosmosis.

One possible explanation for the almost constant velocity as a
function of frequency could be that the amount of charge in the
Debye layer is controlled by a faradaic electrode reaction rather
than by the ohmic current running through the bulk electrolyte. Our
impedance measurement clearly shows that the electrode reaction is
negligible at $f= 1$~kHz and $V_\textrm{rms}= 10$~mV bias, but since
the reaction rate grows exponentially with voltage in an Arrhenius
type dependence, it may still play a role at $V_\textrm{rms}=
0.8$~V. However, previous theoretical investigations have shown that
faradaic electrode reactions do not lead to reversal of the AC
electroosmotic flow or pumping direction \cite{LHO2006}.

Due to the strong nonlinearity of the electrode reaction and the
asymmetry of the electrode array, there may also be a DC faradaic
current running although we drive the system with a harmonic AC
voltage. In the presence of an intrinsic zeta potential
$\zeta_\textrm{eq}$ on the electrodes and/or the glass substrate
this would give rise to an ordinary DC electroosmotic flow. This
process does not necessarily generate bubbles because the net
reaction products from one electrode can diffuse rapidly across the
narrow electrode gap to the opposite electrode and be consumed by
the reverse reaction.

To investigate to which extent this proposition applies, we used the
weakly nonlinear theoretical model presented in \cite{LHOthesis}.
The model extends the standard model for AC electroosmosis by using
the Gouy--Chapman--Stern model to describe the double layer, and
Butler--Volmer reaction kinetics to model a generic faradaic
electrode reaction \cite{BardFaulkner}. The concentration of the
oxidized and reduced species in the diffusion layer near the
electrodes is modeled by a generalization of the Warburg impedance,
while the bulk concentration is assumed uniform, see
Ref.~\cite{LHOthesis} for details.

The model parameters are chosen in accordance with the result of the
impedance analysis, i.e., $C_\textrm{s}= 1.8$~F/m$^2$,
$R_\textrm{ct}= 1$~M$\Omega$, $\zeta_\textrm{eq}= 160$~mV, as
discussed in Sec.~\ref{sec:discussion}A. Further we assume an
intrinsic zeta potential of $\zeta_\textrm{eq}= -100$~mV on the
borosilicate glass walls \cite{Kirby2004}, and choose (arbitrarily)
an equilibrium bulk concentration of 0.02~mM for both the oxidized
and the reduced species in the electrode reaction, which is much
less than the KCl electrolyte concentration of $c= 0.4$~mM.

The result of the model calculation is shown in
Fig.~\ref{fig:Reverseflow}(b). At 1~kHz the fluid motion is
dominated by AC electroosmosis which is solely in the forward
direction. However, at 12.5~kHz the AC electroosmosis is much weaker
and the model predicts a (small) reverse flow due to the DC
electroosmosis for $V_\textrm{rms}< 1$~V.

Fig.~\ref{fig:Reverseflow}(b) shows that the frequency interval with
reverse flow is only from 30~kHz down to 10~kHz, while the measured
velocities remain negative down to at least 1~kHz. The figure also
shows results obtained with a lower Stern layer capacitance
$C_\textrm{s}= 0.43$~F/m$^2$ in the model, which turns out to
enhance the reverse flow.

In both cases, the reverse flow predicted by the theoretical model
is weaker than that observed experimentally and does not show the
almost constant reverse flow profile below 10~kHz. Moreover, the
model is unable to account for the strong concentration dependence
displayed in Fig.~\ref{fig:Reverseflow}(a).

According to Ref.~\cite{Bazant2006a}, steric effects give rise to a
significantly lowered Debye layer capacitance and a potentially
stronger concentration dependence when $\zeta$ exceeds $10\,
k_\textrm{B}T/e \sim 250$~mV, which roughly corresponds to a driving
voltage of $V_\textrm{rms}\sim 0.5$~V. Thus, by disregarding these
effects we overestimate the double layer capacitance slightly in the
calculations of the theoretical flow velocity for $V_\textrm{rms}=
0.8$~V. This seems to fit with the observed tendencies, where
theoretical velocity curves calculated on the basis of a lowered
$C_\textrm{dl}$ better resemble the measured curves.

Finally, it should be noted that several electrode reactions are
possible for the present system. As an example we mention
$2\textrm{H}_2\textrm{O}_{(\textrm{l})} +
\textrm{O}_{2(\textrm{aq})} + 4\textrm{e}^- \rightleftharpoons
4\textrm{OH}^-_{(\textrm{aq})}$. This reaction is limited by the
amount of oxygen present in the solution, which in our experiment is
not controlled. If this reaction were dominating the faradaic charge
transfer, the value of $R_\textrm{ct}$ could change from one
measurement series to another.

\section{Conclusion}\label{sec:conclusion}
We have produced an integrated AC electrokinetic micropump using
MEMS fabrication techniques. The resulting systems are very robust
and may preserve their functionality over years. Due to careful
measurement procedures it has been possible over weeks to reproduce
flow velocities within the inherent uncertainties of the velocity
determination.

An hitherto unobserved reversal of the pumping direction has been
measured in a regime, where the applied voltage is low
($V_\textrm{rms} < 1.5$~V) and the frequency is low ($f< 20$~kHz)
compared to earlier investigated parameter ranges. This reversal
depends on the exact electrolytic concentration and the applied
voltage. The measured velocities are of the order $-5~\mu$m/s to
$-10~\mu$m/s. Previously reported studies of flow measured at the
same parameter combinations show zero velocity in this regime
\cite{Studer2004}. The reason why we are able to detect the flow
reversal is probably our design with a large electrode coverage of
the channel leading to a relative high ratio
$v_\textrm{pois}/v_\textrm{slip}=0.29$.

Finally, we have performed an impedance characterization of the
pumping devices over eight frequency decades. By fitting Bode plots
of the data, the measured impedance spectra compared favorably with
our model using reasonable parameter values.

The trends of our flow velocity measurements are accounted for by a
previously published theoretical model, but the quantitative
agreement is lacking. Most important, the predicted velocities do
not depend on electrolyte concentration, yet the concentration seems
to be one of the causes of our measured flow reversal,
Fig.~\ref{fig:Reverseflow}(a). This shows that there is a need for
further theoretical work on the electro-hydrodynamics of these
systems and in particular on the effects of electrolyte
concentration variation.

\begin{acknowledgments}
We would like to thank Torben Jacobsen, Department of Chemistry
(DTU), for enlightening discussions about electrokinetics and the
interpretation of impedance measurements on electrokinetic systems.
\end{acknowledgments}

\end{document}